\documentclass[epj]{svjour}
\usepackage{epsfig}
\begin{document}
\title{Vitrification in a 2D Ising model with mobile bonds}
\author{Predrag Lazi\'c \and
D. K. Sunko\thanks{email: dks@phy.hr}}
\institute{
Department of Physics,\\Faculty of Science,\\
University of Zagreb,\\
Bijeni\v cka cesta 32, POB 331,\\
HR-10000 Zagreb, Croatia.}
\date{}

\abstract{
A bond-disordered two-dimensional Ising model is used to simulate Kauzmann's
mechanism of vitrification in liquids, by a Glauber Monte Carlo simulation.
The rearrangement of configurations is achieved by allowing impurity bonds to
hop to nearest neighbors at the same rate as the spins flip. For slow
cooling, the theoretical minimum energy configuration is approached,
characterized by an amorphous distribution of locally optimally arranged
impurity bonds. Rapid cooling to low temperatures regularly finds bond
configurations of higher energy, which are both \emph{a priori} rare and
severely restrictive to spin movement, providing a simple realization of
kinetic vitrification. A supercooled liquid regime is also found, and
characterized by a change in sign of the field derivative of the spin-glass
susceptibility at a finite temperature.
}

\maketitle
\section{Introduction}

In a seminal article many years ago, Kauzmann~\cite{Kauzmann48} provided the
now standard~\cite{Angell95} physical picture of vitrification in liquids,
which emphasized that obstruction of configurational rearrangement by
molecular interactions could lead to such an inefficient sampling of phase
space, that thermodynamic equilibrium could not be reached for very long
times. The ability to change configuration practically disappears at the
glass transition, while orientational disorder does not change significantly,
being high and relatively static in both the supercooled liquid and the
glass. For our simple purposes, the scenario may be compared to a log-jam.
There is nothing in the Hamiltonian to prevent orderly stacking, and indeed,
vitreous liquids possess a crystal state of lower free energy. But once the
logs are jammed, it is (very) improbable they will be ordered by random
pushing and pulling.

By contrast, in spin glasses~\cite{Binder86,Young95}, the Hamiltonian itself
forbids crystal order, because of impurities, or lattice topology. Now
disorder in terms of the original microscopic variables becomes an
equilibrium property at all temperatures, leaving open the possibility of a
true thermodynamic phase transition in terms of some other, less-than-obvious
order parameter. In searching for it, the basic concept is frustration,
introduced~\cite{Toulouse77} to distinguish `intrinsically' disordered states
from those where apparent disorder can in fact be removed by redefining
parameters of the Hamiltonian. These `serious' and `trivial' kinds of
disorder may be called `glassy' and `amorphous', respectively. 

In the present article, we bring the two subjects together, to obtain a toy
model of kinetic vitrification. In brief, our algorithm gives a structural
glass in a spin-glass setting. We begin with the discrete bond-disordered
Ising model in two dimensions. It is known~\cite{Bhatt88} that it has no
spin-glass transition at $T>0$. Topologically frustrated configurations are
too rare in two dimensions to contribute significantly to the sample-averaged
free energy, when each random arrangement of impurities contributes a sample
with equal \emph{a priori} probability. We refer to this standard
quenched-disorder picture as `frozen' or `frozen bonds', reserving the word
`quenched' as follows.

The model we consider is obtained from the above by allowing the impurity
bonds to move. Throughout the article, `quenching' means that this system,
with movable bonds, is first equilibrated at high temperature, and the
temperature then brought down to some low value in a single step, after which
the evolution continues at that low temperature. Since the bonds move, there
is no averaging over samples; the complete impurity configuration space is
available to a single sample. Of course, the efficiency with which this space
is explored after the quench critically depends on the nature of bond
movement. We focus on the case where they move locally, \emph{i.e.} a bond
can only exchange places with another bond, impinging on the same site. Our
main point is that such an algorithm, which would be very inconvenient to
study the ground state of the disordered Ising model, because it is easily
trapped in metastable states, is very convenient from the structural-glass
point of view, because the metastable states in which it does get trapped are
quite exceptional among all possible impurity arrangements, and are analogous
to vitreous states in the structural sense.

It is easy to imagine how this happens: the bonds keep moving until they get
stuck. These `stuck' configurations are precisely the vitreous ones, which
are (by definition) those in which evolution appears to cease before the
ground state is reached. Various signals of their exceptional nature are
studied throughout the article; let us only mention that their energy is
markedly lower than the average energy in the frozen scenario, in fact close
to the theoretical ground state energy (which can also be reached, by slow
cooling). Nevertheless, it is finitely above the ground state, making them
metastable.

In the present work, we consider only the fast evolution after a quench,
leading to the above-mentioned metastable states, and the evolution of these
states at intermediate times, but not their relaxation to the ground state,
occuring at much longer time scales. The ground state (annealed system) is a
comparatively trivial amorphous configuration, and our only interest in it is
that it establishes the formal metastability of the long-lived quenched
states. In this, of course, our toy model is different from a true vitreous
liquid, where the ground state is an ordered crystal. The model analogue of
melting involves no order-disorder transition.

In the description of our results, we use established language (\emph{e.g.},
melting) and theoretical tools (\emph{e.g.}, the spin-glass susceptibility)
somewhat outside their original context. The model is purely kinetic, without
a thermodynamic phase transition, so the interesting (vitreous) behaviour
appears in quasi-equilibrium, established at the above-mentioned intermediate
time scale. We are guided by an intuitive analogy between spins and bonds in
the disordered Ising model, on the one side, and the orientational and
configurational degrees of freedom, respectively, of a glass-former, on the
other.

One class of Ising-like models, often used to study vitrification, are
constrained kinetics models, such as the Frederickson-Andersen
model~\cite{Frederickson84}. They have no intrinsic disorder, but frustration
is introduced by dynamical restrictions on changing the spin states, which
take the part of configurational variables. Among work on variants of the
disordered Ising model, our program is closest to that of the frustrated
percolation problem~\cite{Coniglio97,Cataudella96}. However, that approach
views impurity bonds as obstacles to rearrangement, playing a role similar to
the dynamical restrictions, mentioned above. In our work, bonds are imagined
to represent the configurational degree of freedom itself, so their movement
becomes of physical interest. It is responsible for the appearance of
metastable states, the `supercooled bond liquid', which have a distinct
physical characteristic, a first-order zero in the field derivative of the
spin glass susceptibility, not shared by the states on the cooling curve
connected to the ground-state manifold. This liquid, and the accompanying
glass, are the main subject of the present article.

\section{Model and algorithm}\label{mod}

The model is the bond-disordered two-dimensional Ising model on a square
lattice~\cite{Edwards75}:
\begin{equation}
H=-\sum_{\left<ij\right>}J_{ij}S_iS_j-B\sum_iS_i\equiv
-\sum_{\left<ij\right>}J_{ij}S_iS_j-B\cdot M,
\label{ham}
\end{equation}
where the `spins' take the values $S_i=\pm 1$, as well the bonds, $J_{ij}=\pm
1$. Here $M$ is the total magnetization, calculated as above, not an
independent parameter. The fraction of antiferromagnetic (negative $J_{ij}$)
bonds is a fixed number $p$; in this paper, $p=70$\%. The key new feature is
the ability of bonds to move; this is considered to be purely diffusive, so
there is no kinetic contribution to the Hamiltonian.

Spin updates proceed by flipping a spin at random; the criterion for
acceptance is Glauber's. Between spin updates are bond updates; the test move
is to exchange two bonds of the opposite sign, impinging on the same site.
This is accepted or rejected, again by the Glauber criterion. Which bonds are
exchanged, and on which site, is chosen at random, taking care, of course,
that the two bonds involved are indeed of the opposite sign. There is no
regular `sweeping' of the lattice, to avoid artificial correlations between
spin and bond updates.

There still remains a physical freedom in the update protocol. One might
consider, in general, flipping $s$ spins and moving $b$ bonds, before
submitting the new configuration for acceptance. These are large steps in the
random-walker picture. They are unlikely to be accepted in general, but once
the system is trapped in a glassy state, such steps may significantly affect
its chance to get out. These rare `cooperative' fluctuations are considered
responsible for re-crystallization of vitreous liquids in the Kauzmann
picture, sometimes after `archeologically' long times, as in window glass.

In this work, we only focus on driving the system into a glassy state, not
getting it out. On this `laboratory' time scale, alternating one-spin and
one-bond updates, each subject to acceptance in turn, provide the dominant
quasi-equilibration mechanism. The very-long-time behavior of the model is
the subject of a limited investigation in Section~\ref{equil} below, but
still neglecting multistep (cooperative) processes, which could be important
for the asymptotic regime. Parenthetically, we note that mode-coupling theory
introduces an artificial thermodynamic transition, and therefore certainly
has the wrong asymptotics, but even this does not affect the intermediate
time scales, for which it was devised~\cite{Schmitz93}.

We consider two situations: annealed, where the system is equilibrated at
high temperature ($T=5$, but any $T>2$ will do), and then slowly cooled (in
steps of $0.05$, waiting to equilibrate between temperature steps); and
quenched, where it is brought down to the desired temperature in a single
step from the high-temperature equilibrium, and then quasi-equilibrated. The
high-temperature starting point is generated anew for each quenched point, so
the agreement of annealed and quenched curves for not-too-low temperatures is
also a check of equilibration. Similarly, the fact, that the annealed curve
nearly reaches the ideal (defect-free) ground state energy at $T=0.05$, our
lowest temperature, is indication that the cooling rate for annealment
($\Delta T\sim 10^{-9}$ per MC step) is sufficiently slow for our purposes.

For comparison, we also simulate some systems with frozen bonds. This opens
the question of sample-averaging. We find that self-averaging is sufficient
for our limited purpose, since the samples involved are rather large
($100\times 100$ and larger). To be safe, we have made some runs, averaging
batches of up to $500$ small ($\sim 10\times 10$) samples, and indeed no
claims made in the paper, based on the large self-averaged ones, were
affected. Note, also, that the frozen bond arrangement at high temperature is
different for each cooling run, so the above-mentioned check of equilibration
is also a check of self-averaging in the frozen case.

\section{Main result}\label{main}

All simulations were performed in a uniform external field, whose non-trivial
range is $0\le B\le 4$. Then the theoretical minimum energy per site of the
annealed system with a fraction $p$ of antiferromagnetic bonds is
\begin{equation}
U_{\mathrm{min}}=-2-(1-p)B,
\label{umin}
\end{equation}
which assumes the absence of defects: each spin, turned opposite to the
field, is connected to its neighboring sites by antiferromagnetic bonds,
which are all used up in this way. These `crosses' carry no frustration,
obviously: a simultaneous sign change of the spin and its four bonds leads to
a perfect ferromagnet. There is a ground-state entropy, associated with their
random placing on the lattice, so this ideal state is amorphous. It is
our model's analogue of the crystal state of vitreous liquids.

A convenient robust measure of additional disorder is the energy mismatch
between this reference state and one obtained in an actual run. This can be
extended to any temperature by using the reference system's cooling curves,
once it is checked that its zero-temperature limit is near the theoretical
amorphous one, \emph{i.e.}, that it is sufficiently annealed.

\begin{figure}[tb]
\center{\epsfig{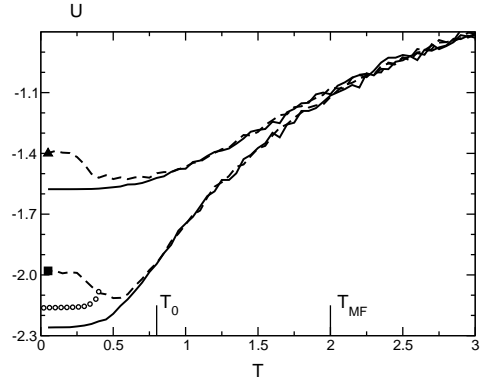}}
\caption{Energy per site for $p=0.7$ and $B=1$. Full curves: annealed. Dashed
curves: suddenly cooled. Lower (upper) two curves: with (without) bond
movement. Open circles: quench to $T=0.4$ and slow cooling. Symbols at
$T=0.05$: see Figs.~\ref{quench} and~\ref{freeze}. $U_{\mathrm{min}}=-2.3$
[Eq.~(\ref{umin})], lattice $100\times 100$.}
\label{ener}
\end{figure}

Figure~\ref{ener} shows the energies per site for a single cooling run, in an
external field $B=1$ and $p=0.7$. Four curves are given, annealed and
suddenly cooled, with and without bond movement allowed. Two energy scales
appear: a higher one, where the curves with bond movement separate, and a
lower one, distinguishing annealed and suddenly cooled systems. Unexpectedly,
the higher scale is about the square root of the coordination, which is the
mean-field transition temperature of the model (\ref{ham}),
$T_{MF}=\sqrt{z}=2$; this may be a coincidence. The lower one, denoted $T_0$,
is the model's melting temperature, a notion made precise in
section~\ref{slt} below. The main interest in this work is the range beneath
it.

Below $T_0$, the dashed curves rise with falling temperature, because each
point on them is generated anew by a drop from high temperature. The lower
the temperature to which they are quenched, the `worse' (energetically) the
state in which they end. In the next section, we shall discuss a third type
of cooling run, where the system is cooled slowly after a quench from high
temperature. An example of this is also given in the figure, showing the
thermodynamically expected behavior.

\begin{figure}[tb]
\center{
\epsfig{file=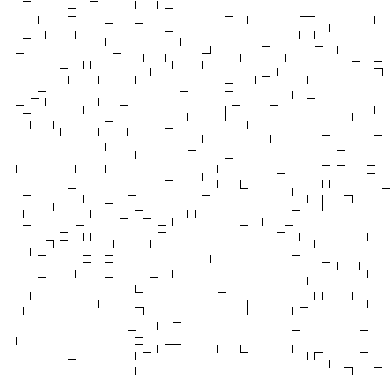,height=50mm}
}
\caption{A $50\times 50$ segment of the quenched state, marked by a filled
square in Fig.~\ref{ener}. Only uncompensated bonds are shown. The
preponderance of singletons indicates nearly all the disorder is glassy.}
\label{quench}
\end{figure}

\begin{figure}[tb]
\center{\epsfig{file=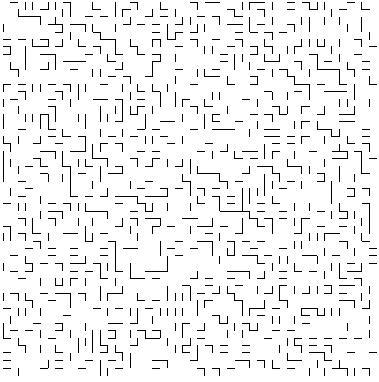,height=50mm}}
\caption{A $50\times 50$ segment of the frozen, rapidly cooled state, marked
by a filled triangle in Fig.~\ref{ener}. Only uncompensated bonds are shown.
Much of the disorder is amorphous.}
\label{freeze}
\end{figure}

While the two dashed curves look similar, they refer to very different
states. In Figures~\ref{quench} and~\ref{freeze}, we draw the respective
lattices at low temperature, presenting only uncompensated bonds, \emph{i.e.}
those which are of the `wrong' sign relative to the spins they connect. The
exact ground states have as few of these as possible (a nonzero number may be
topologically unavoidable~\cite{Vogel98}). The quenched state obviously has
more, but the point is that they nearly always appear as well-separated
singletons, carrying topological frustration. In other words, the frustration
of our vitreous states is near the maximal possible one, for a given energy
difference from the ground state. In consequence, spin orientation is fixed
by a `majority vote' of bonds \emph{at every single site} (almost). The
comparatively small energy difference between the ground and vitreous states
is misleading; the two are separated by a topological barrier. The fact that
such \emph{a priori} rare states can be generated with so little effort,
about 50 updates per site, is an important practical point of this work.

By contrast, in the case with frozen bonds, there appear many sites with two
compensated and two uncompensated bonds, making the spin orientation
energetically indifferent. Many of the corresponding plaquettes are not
frustrated, so the energy difference relative to the ground state is largely
due to trivial (amorphous) disorder. Because of this, the magnetization of
the frozen-bond system is a sum of fluctuating contributions even at low
temperature, making it constant in time only on the average, while the
magnetization of the vitreous (quenched) system is a sum of spins which are
individually fixed in time for the lifetime of the glass, as would be the
case for all times with a perfect Ising ferromagnet at sufficiently low
temperature.

\begin{figure}[tb]
\center{\epsfig{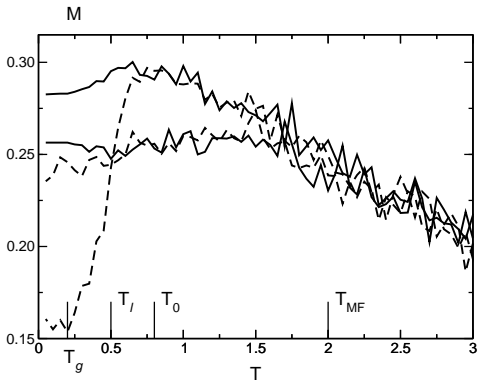}}
\caption{Magnetization per site for $p=0.7$ and $B=1$. Full curves: annealed.
Dashed curves: quenched. Middle two curves: frozen bonds.
The other two curves: mobile bonds. $M_{\mathrm{max}}=0.3$
[Eq.~(\ref{umin})], lattice $100\times 100$.}
\label{magn}
\end{figure}

Figure~\ref{magn} shows the magnetization. For a frozen bond configuration,
it barely distinguishes between annealing and quenching. What is unexpected
is the difference between the annealed and quenched magnetizations, when bond
movement is allowed. The annealed magnetization at $T\approx 0$ is higher
than when bonds are frozen, reasonably enough. The quenched one, however, is
much lower, indicating a system in which spin orientation is controlled by
an internal field. It is very sensitive to the $T_0$ scale of
Fig.~\ref{ener}, while the same scale is hardly visible in the other three
curves.

It remains to show that the state quenched (in our sense, with mobile bonds)
below $T_g\approx 0.2$ fulfills the criterion for a spin glass. We call it a
`bond glass', to emphasize the role of bond kinetics in its formation; it is
not in thermodynamic equilibrium. The transition region $T_g\approx
0.2<T<0.8\approx T_0$ in Figure~\ref{magn} is then a `supercooled bond
liquid', characterized by a scale $T_\ell\approx 0.5$. For temperatures above
$T_0$, we have a bond liquid. Carefull cooling below the melting temperature
$T_0$ leads to the amorphous solid, described above. (Note the small downturn
below $T_0$ in the slowly cooled curve, which seemed headed towards the
theoretical maximum magnetization $M=0.3$. It is due to defects in the
amorphous structure. The unavoidability of these is the subject of
bond-percolation studies~\cite{Vogel98}.)

We consider the spin-glass susceptibility, which is a time-asymptotic
four-spin correlation function~\cite{Young95,Bhatt88}:
\begin{equation}
\chi_{\mathrm{SG}}=V\int_{-1}^1q^2P(q)dq,
\label{defchi}
\end{equation}
where
the `volume' $V$ is the number of spins in the lattice, and $P(q)$ is the
distribution of the quantity
\begin{equation}
Q=\lim_{\tau\to\infty}{1\over\Delta\tau}\sum_{t=1}^{\Delta\tau}
\left[{1\over V}\sum_{i=1}^VS_i^{(1)}(\tau+t)S_i^{(2)}(\tau+t)\right],
\label{Q}
\end{equation}
where the superscripts on the spins denote two copies of the system,
independently evolved, and `infinity' is some intermediate time scale, before
ergodicity is re-established, as discussed below. For each temperature, we
first quasi-equilibrate the system, as to produce a point in the first two
figures, then make two copies of the lattice, identical in both spins and
bonds, so an artificial divergence is introduced in $\chi_{SG}$ at
$\tau=t=0$. This is allowed to decay by evolving the copies independently for
a further $\tau=5\times 10^6$ MC steps, each consisting of a spin and a bond
trial, or only a spin trial, for the cases with frozen bonds. The
distribution $P(q)$ is obtained by considering the next $\Delta\tau=2.2\times
10^6$ steps.

In the frozen case, this is done differently~\cite{Bhatt88}: one randomizes
the spins in both copies of the lattice, with the \emph{same} bond
arrangement, and waits for the correlation function to build itself up again
from zero. We cannot do that, because the bonds also move: once the spins are
randomized, bonds start flowing, and the lattice copies end up as distinct
samples, as soon as the vitreous state is re-established, which takes a very
short `$\beta$-relaxation' time, only $\sim 50$ updates per site in our
algorithm. In the usual picture of a glass free energy landscape, the two
copies end in distinct deep and narrow minima, and the scale associated with
tunneling between such minima is the long `$\alpha$-relaxation' time. In the
same language, the time window $\Delta\tau$ is opened after `infinity' $\sim
20$ times the $\beta$-relaxation period, \emph{i.e.} in the
quasi-equilibrated regime, $\tau_\beta<\tau<\tau_\alpha$ for all $T$, which
is the regime of interest.

The use of the spin-glass susceptibility should not be construed to mean,
that there is a true thermodynamic transition in the model. As long as the
bonds move, all the reported phenomena are kinetic in origin. Asymptotically,
$\chi_{SG}$ is zero at all temperatures, but the time needed to see this
increases exponentially with falling temperature, as discussed in the next
section. 

\begin{figure}[tb]
\center{\epsfig{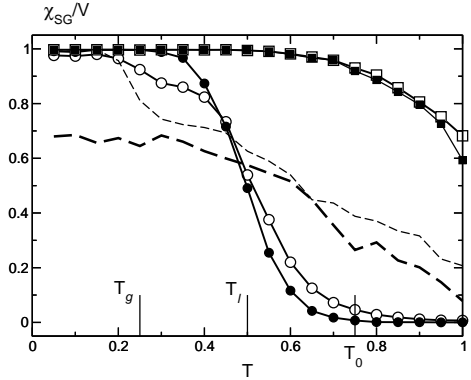}}
\caption{Normalized spin-glass susceptibility on a $120\times 120$ lattice,
for a number of quenching scenarios. Empty symbols: $B=1$. Filled symbols:
$B=0^+$. Circles: quench with movable bonds. Squares: same as circles, but no
bond movement after $\tau=t=0$. Thick (thin) dashed line: $B=0^+$ ($B=1$),
random frozen bonds. Zero-field $T_g$, $T_\ell$ and $T_0$ are also given.}
\label{chi}
\end{figure}

The results are shown in Figure~\ref{chi}, which we discuss for the remainder
of this section. The open circles correspond to the quenched bonds in
Figure~\ref{magn}. The range $T<0.2$, where the magnetization saturates at
$M\approx 0.15$, is clearly linked to a divergence in $\chi_{SG}$. The
crossover region, $0.2<T<0.8$, is also evident in both figures.

The thin dashed line corresponds to the frozen-bond scenario, whose
magnetization is given by the nearly horizontal dashed line in
Figure~\ref{magn}. It saturates at $\chi_{SG}/V=1$ as well, which may cause
doubt about the alleged special nature of the state, found by bond movement.
To dispel it, we repeat all of the above with the magnetic field turned off.
(We put $B=0.001$ to break inversion symmetry, and denote this $B=0^+$.) The
frozen-bond susceptibility then drops to the lower, thick dashed curve, which
no longer shows saturation. This indicates its divergence at finite field was
in fact trivial, due to the difficulty of flipping spins when $T\ll B$, so
the spins on the two copies `marched together' for a long time.

The quenched susceptibility shows precisely the opposite behavior, given by
the filled circles. For $B=0^+$ it rises above the $B=1$ points, to achieve
nearly perfect divergence for $T<0.3$, but drops below them in the
high-temperature part, so the net effect is a narrower crossover (supercooled
liquid) region, $T_g=0.25<T<0.75=T_0$, its center (half-point of
$\chi_{SG}/V$) remaining at $T_\ell=0.5$. The low-temperature divergence in
the quenched case is due to the bonds `holding' the spins, in place of any
external field. The field \emph{widens} the supercooled liquid regime, and
produces a less perfect glass (slightly smaller $\chi_{SG}$ below $T=0.2$).
At $B=0^+$ nothing is visible in the magnetization, which fluctuates around
zero at all temperatures.

We now turn to the role of bond rearrangement in the supercooled liquid. We
restart the evolution from the same copies of the lattice as gave the circles
in Figure~\ref{chi}, but now disallow further bond movement. In other words,
we proceed as with frozen bonds (the dashed curves), but for the special bond
configurations found by quenching (Fig.~\ref{quench}), not random ones
(Fig.~\ref{freeze}). Then the open and filled circles go over into the
respective squares. The supercooled liquid has disappeared, replaced by a
glass up to $T\approx T_\ell$. This is in accord with the idea, that changing
configurations is essential for the liquid. It may appear that the crossover
has merely moved to higher temperatures, not shown in the figure. But $T=0.8$
is already the melting temperature in Figure~\ref{magn}, so the system cannot
be `supercooled' at $T>T_0$. It melts, as it eventually must, by a trivial
high-temperature degradation of the initial spin orientations. However, the
rarity of the bond configurations involved is evident in the lack of
field-dependence of the susceptibility (the squares nearly overlap), in sharp
contrast with the behavior of the dashed curves.

\section{Equilibration}\label{equil}

The basic point of this article is that the algorithm with bond movement is
physically interesting in itself, as a model realization of a structural
glass. At present, we have no physical argument that the degrees of freedom
of any particular real glass map onto our model. So one has at best an
analogy, or `toy model', and the idea is to use the analogy to guide model
investigations. Some quantitative evidence will be presented in this section,
that the properties uncovered in this way are not spurious.

\subsection{Short and long times}\label{slt}

\begin{figure}[tb]
\center{
\epsfig{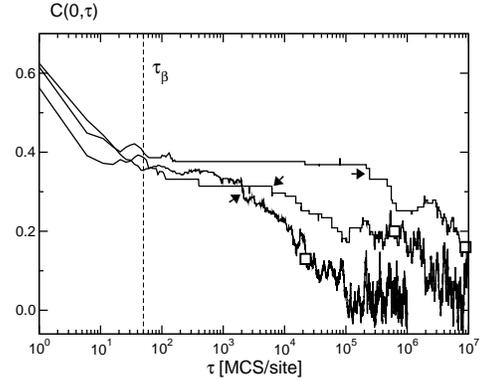}
}
\caption{Spin autocorrelation function, quenched case, high-temperature
initial state. Left to right by arrows: $T=0.4$ ($L=25$), $T=0.3$ ($L=15$),
$T=0.25$ ($L=15$). Arrows: dip marking $\tau_{\alpha 1}$. White squares:
points with abscissa given by Eq.~(\ref{ocjena}). Dashed line: a
minimal $\tau_{\beta}$, for all cases.}
\label{decay}
\end{figure}

In Figure~\ref{decay}, we show the spin autocorrelation function
\begin{equation}
C(t,\tau)={1\over V}\sum_{i=1}^V\left<S_i(t)S_i(t+\tau)\right>,
\label{auto}
\end{equation}
where the brackets denote an average over time-windows of 10 MCS/site. The
reference state ($t=0$) is taken to be the high-temperature equilibrium
($T=5$) before the quench. A typical two-time evolution is observed. After a
short $\beta$-relaxation time $\tau_\beta$, 50--100 MCS/site, the system
enters a long-lived low-temperature state. This intermediate period, in which
evolution almost seems to stop, ends at  a time $\tau_{\alpha 1}$, when
$C(0,\tau)$ begins to decay further, reaching zero only after a much longer
time $\tau_{\alpha 2}$. The latter is usually called $\tau_\alpha$, the scale
at which ergodicity is re-established. This is the time needed to tunnel
between the `deep and narrow' minima characterizing the glass free energy
landscape. In the same picture, $\tau_{\alpha 1}$ would be the time needed to
exit the original minimum, which trapped the system at $t\approx\tau_\beta$.

It is easy to estimate $\tau_{\alpha 2}$ in the present model, because
Fig.~\ref{quench} indicates maximal topological frustration is a reasonable
hypothesis. This implies, roughly, that \emph{each} spin needs to flip
against the local field before the state is completely dissolved. In other
words, the reciprocal probability for a single spin to flip is the waiting
time for dissolution, measured in updates per site. The energy difference for
a flip is $\Delta E=4$ or $\Delta E=8$, depending on whether the spin is held
in place by three or four compensated bonds. Taking the lesser of these gives
an underestimate,
\begin{equation} \exp\left({4\over T}\right)<\tau_{\alpha
2} \mbox{ }\mathrm{[MCS/site]},
\label{ocjena}
\end{equation}
which we find to be logarithmically correct (Fig.~\ref{decay}). It predicts
$\tau_{\alpha 2}>5\times 10^{34}$ updates per site at $T=0.05$.

The success of the simple estimate (\ref{ocjena}) indicates that the
algorithm with nearest-neighbor bond diffusion realizes, more or less, the
maximal time scales possible in the model, for a purely kinetic scenario. But
it also implies that the scenario is indeed kinetic: nothing really happens
at $T_g$, except that a time scale becomes too long to follow. This is the
reason the present approach is computationally so cheap. One knows in advance
that, say, $\chi_{SG}$ in Fig.~\ref{chi} is asymptotically zero. The question
is, for what delay [$\tau$ in Eq.~(\ref{Q})] it is most informative of the
processes in the system. This turns out to be fairly modest, $\sim 10^3$
updates per site. By then, the value of $T_g$ is as well established as for
$10^5$ updates, while the $\tau_\alpha$ for most of the higher temperatures
are still not reached, giving the largest possible range in temperature, for
which $0<\chi_{SG}<1$. Of course, there is a highest temperature, where
$\tau_{\alpha 2}\to\tau_\beta\approx 50$. This is just the melting
temperature $T_0$: $\exp(4/T_0)\approx 50$ gives $T_0\approx 1$, as found in
Fig.~\ref{ener}. Above it, $\chi_{SG}$ is zero at all times.

\subsection{Aging}

\begin{figure}[tb]
\center{
\epsfig{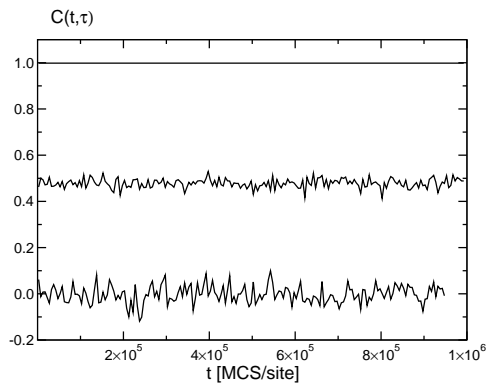}
}
\caption{Spin autocorrelation function, quenched case, for $t>\tau_\beta$.
Bottom to top: 1)~$T=0.6$, $L=25$, $\tau=10^5$ MCS/site; 2)~$T=0.6$,  $L=15$,
$\tau=10^3$ MCS/site; 3)~$T=0.2$, both cases. Time-averaging is over $10^4$
updates per site.}
\label{correl}
\end{figure}

After the initial relaxation has passed, the supercooled liquid/glass spends
a long time in a state which is nominally metastable, since its energy is
above that of the annealed state (Fig.~\ref{ener}). However, it is
equilibrated from the point of view of correlations; there is no
`aging'~\cite{Kisker96} in the metastable state.  In such a
quasi-equilibrium, the autocorrelation function (\ref{auto}) does not depend
on choice of origin $t$, apart from noise, as shown in Figure~\ref{correl}.
Here all $t>\tau_\beta$, so the initial (reference) state is always a
quasi-equilibrated one. The amplitude of the noise generally increases with
$\tau$ and the temperature, while it decreases as the system gets larger. At
$T=0.2$, $C(t,\tau)=1$, and there is no noise, for the trivial reason, that
the state below $T_g$ does not evolve at the scale of the waiting times
exhibited here.

\subsection{Rattling and relaxation}

It is illustrative of the qualitative behavior of the model, to look at the
acceptance rates of various kinds of trial moves. They are the inverses of
the time scales involved. We introduce a third type of cooling run, most
similar to a real experiment. In zero field, the system is first quenched
into the supercooled region (below $T_0$), and then slowly cooled through
$T_g$, in steps of $0.02$. We look at the time evolution leading to
quasi-equilibration (not after it, as in the study of $\chi_{SG}$). After the
temperature is brought to the desired value, a $120\times 120$ lattice is
subject to $7.2\times 10^6$ spin trials, interlaced with the same number of
bond trials. The principal quantity monitored is the number of accepted
relaxation steps, \emph{i.e.} those in which the energy changes either way
(positive changes are counted as relaxation, because they also mean the
system is `working').

\begin{figure}[tb]
\center{\epsfig{file=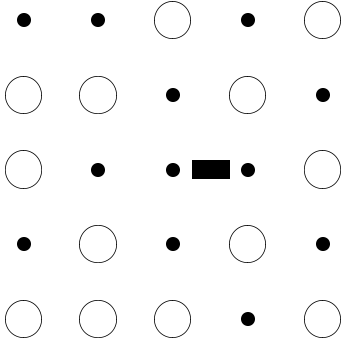,height=50mm}}
\caption{A `caged' AF bond, free to rotate around the central spin. The
other bonds (not shown) are all of the correct sign, corresponding to the
spins they connect. This bond could be freed by flipping any of the four
spins in the centers of the square sides, except the right-hand one.}
\label{bond}
\end{figure}

The estimate (\ref{ocjena}) indicates that by lowering the temperature below
$T_g$, one can really prevent the system from reaching the annealed state in
any fixed time. In addition to the disappearance of spin flips, the nature of
bond movement changes below $T_g$, and only `rattling'~\cite{Angell95} of the
kind depicted in Figure~\ref{bond} survives. For our purposes, we do not
define `rattling' geometrically, but by the property that the move is
energetically indifferent, so that it cannot contribute to annealing, by
definition.

\begin{figure}[tb]
\center{\epsfig{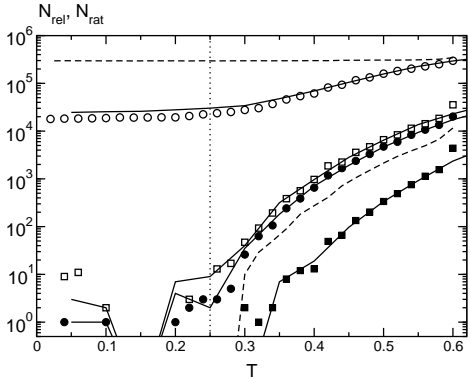}}
\caption{Number of accepted steps during quasi-equilibration of a large
system. Symbols: slow cooling after quench to $T=0.6$; circles: bond moves;
squares: spin flips; filled: relaxation; empty: rattling. Full lines: same as
symbols, but annealed. Dashed lines: spin moves with frozen bonds, upper:
rattling, lower: relaxation. Vertical dotted line: $T_g$ from
Fig.~\ref{chi}.}
\label{nrel}
\end{figure}

In Figure~\ref{nrel}, we show the total number of relaxation steps, accepted
during the run above. Bond relaxation disappears around $T_g$, while spin
relaxation stops earlier, and is a rough order of magnitude less significant
everywhere. Spin rattling vanishes along with bond relaxation. Thus both the
energy and magnetization cease to evolve below $T_g$. This is also visible in
Fig.~\ref{ener}, the open circles there becoming strictly horizontal below
$T_g$. Bond rattling, on the other hand, remains high and relatively constant
throughout, in keeping with the analogy to `caged' configurational movement.

This is in marked contrast to the behavior of a quen\-ched, then slowly
cooled system with random frozen bonds, shown by dashed lines. Spin moves in
this case behave roughly as bond moves above. Relaxation stops above $T_g$,
but spin rattling is remarkably constant down to lowest temperatures, as one
would expect from the absence of a spin glass in 2D, \emph{i.e.}, of an
internal field to hold the spins in place. On the other hand, spin rattling
in the 3D Edwards-Anderson model persists even below the spin-glasss
transition. This is in contrast to the 2D vitreous state, in which there are
virtually no loose spins at all (Fig.~\ref{quench}).

Finally, we compare the results of a standard annealing run, shown by full
lines. They are in excellent quantitative agreement with the supercooled
liquid/glass --- despite the latter being by definition metastable, as it
consists of higher-energy configurations. This is further illustration of the
long-lived nature of the metastable quasi-equilibrium. In the absence of
frustration, the difference in energy would translate into a larger number of
accepted steps in the metastable case. This is actually observed at $T=0.6$,
but only because the system is quenched from $T=5$ there. Even when shedding
such a large energy, the excess steps are almost exclusively due to spins,
indicating that the bond (`configurational') degree of freedom is most
affected by frustration.

\section{Discussion}\label{dis}

The main problem with the identification of the glass in Section~\ref{main}
is that $\chi_{SG}$ is not by itself a measure of frustration. It may
function as such in a spin-glass setting, where one averages over many random
bond configurations, so if two lattice copies `march in lockstep' on the
average, one can safely assume this is not due to the influence of the
occassional ordered configuration.

Not so in our case, where the whole idea is that mobile bonds find rare
configurations. In particular, they find the optimal amorphous ones, which
show the same persistence of the initial divergence in $\chi_{SG}$, as would
a perfect ferromagnet. The glassy configurations are not far in energy
(Fig.~\ref{ener}), and they, too, contain large admixtures of the optimal
arrangement (Fig.~\ref{quench}).

\begin{figure}[tb]
\center{\epsfig{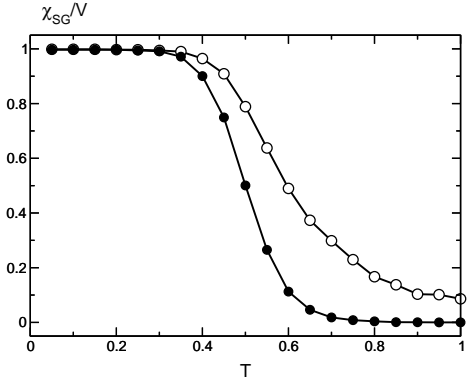}}
\caption{Same as Figure~\ref{chi}, but for the annealed state. Note the
different field dependence.}
\label{amorph}
\end{figure}

The problem reduces to this: can one find a qualitative criterion to
distinguish the `amorphous solid with defects', the uppermost of the
magnetization curves in Fig.~\ref{magn}, from the `supercooled liquid/glass',
the curve which forks down from it below $T_0$? Luckily, yes. It is provided
by the field-dependence of $\chi_{SG}$, shown in Figure~\ref{amorph} for the
amorphous solid. The filled circles ($B=0^+$) show the same behavior as in
Figure~\ref{chi}, in line with the doubts expressed above. But the $B=1$
curve (empty circles) is now entirely above the other one; in the annealed
case, turning on the field increases $\chi_{SG}$ at all temperatures. The
criterion follows, that if the field derivative of the spin-glass
susceptibility
\begin{equation}
\left.{\partial\chi_{SG}(T)\over\partial B}\right|_{B=0^+}
\label{third}
\end{equation}
is positive at all temperatures, we have a solid. If it is negative around
zero and changes sign at a finite $T=T_\ell$, we have a supercooled liquid.

This criterion makes physical sense. In the amorphous solid, the bonds are as
good as frozen, since any rattling occurs in an optimally arranged background
of spins and bonds. The spins opposing the field are in a minority (witness a
finite magnetization), so increasing the field will make the system more
rigid, increasing $\chi_{SG}$ at all temperatures.

Not so in the supercooled liquid, where the bonds rattle against a suboptimal
background. At low temperatures, spin flips induced by the field will
increase the range of bond movement, which in turn has a cascading influence
(stochastic, of course) on spins in the bond path, degrading the correlation
between the two copies of the lattice (see Fig.~\ref{bond}). At higher
temperatures, however, there are already many spin flips, so this effect is
`played out'; the spin majority orchestrated by the field then increases the
correlation, like in the solid. The temperature at which the two tendencies
balance is just $T_\ell$.

This does not mean that $T_\ell$ is a thermodynamical (time-independent)
temperature, indeed $T_g$ itself is not, in any kinetic model. Given that it
refers to the liquid state,  $T_\ell$ is subject to the same time evolution
as $\chi_{SG}$ in that regime, or the correlation function in
Fig.~\ref{decay}. Nevertheless, the above argument implies that the
appearance of a point, where the zero-field and finite-field susceptibilities
cross, is generic for the supercooled liquid in this model, and will always
be present for times between $\tau_\beta$ and $\tau_\alpha$. Since
$\tau_\beta$ is so short, it can be observed without large numerical effort.

There are two things to note. First, the criterion for a supercooled bond
liquid implies (mildly) paradoxical behavior: the bonds `waiting to break
out' make the susceptibility \emph{decrease} in response to an applied field,
when $T<T_\ell$. The field has a net disorganizing effect on the spins,
because it probes the configurational metastability of the supercooled
liquid. Put simply, the  uniform external field is in competition with the
stochastic internal field, holding the spins in position for $T<T_\ell$.

Second, the glass itself is defined only quantitatively, as the manifold of
states found on that part of the cooling curve of the supercooled liquid,
where $\chi_{SG}/V$ saturates at unity. A qualitative characterization of
$T_g$ is not to be expected, on general grounds. In this context, the switch
from circles to squares in Figure~\ref{chi} may be interpreted, that if one
takes a `snapshot' of the system below $T_\ell$ at an arbitrary point in
time, no analysis of that one snapshot can tell, whether it was taken above
or below $T_g$.

It is easily shown that the maximally frustrated (vitreous) bond
arrangements, found by the quenching algorithm, are a very small subset of
all possible ones. For a state to be maximally frustrated, uncompensated
bonds must be singletons, like in Fig.~\ref{quench}, each well separated
from the others by a `layer' of compensated ones, effectively `using up'
$q>1$ bond positions for each singleton. Thus uncompensated bonds can choose
only among some $2V/q$ positions, less than all $2V$ available ones. The
number of distinct states is largest when the number of uncompensated bonds
is one-half the total number of available positions, giving at most
\begin{equation}
{2V/q\choose V/q}\sim 4^{V/q}
\label{single}
\end{equation}
states of this kind. On the other hand, supposing for simplicity that one has
an equal number of positive and negative bonds ($p=0.5$), the number of ways
to place them on the lattice without restriction is
\begin{equation}
{2V\choose V}\sim 4^{V},
\end{equation}
of which the above $q$-th root is a vanishingly small fraction, when $V$ is
large. Energetically, the vitreous states are well separated from the bulk of
the arbitrary ones, witness Fig.~\ref{ener}. They correspond to the `deep and
narrow' minima of glass theory. In fact, since their energy is determined by
the number of singletons, Eq.~\ref{single} is also their degeneracy,
`exponentially large', as the general picture assumes. Since the algorithm is
trapped in only one of them during any given run (Fig.~\ref{decay}), one has
`breaking of ergodicity'~\cite{Angell95} for times less than $\tau_\alpha$.

Let us summarize the analogies between our toy model and vitreous liquids.
Bond and spin movement correspond to the configurational and orientational
degrees of freedom, respectively. The diffusivity of bond movement reflects
the absence of a kinetic energy contribution to the glass transition. The
supercooled bond liquid turns into a glass if configurational rearrangements
are artificially turned off. Spin disorder changes very little across the
glass transition, but spin movement ceases there, as well as all relaxation.
Configurational (bond) rattling persists to lowest temperatures. Somewhat
above our original ambitions, the zero-field glass transition in the model
occurs at about one-third the melting temperature, $T_g\approx T_0/3$, an old
rule of thumb in glass formers~\cite{Kauzmann48}. 

\section{Conclusion}
 
We have realized a structural spin glass in the two dimen\-sional
bond-dis\-ordered Ising model on a square lattice, by taking into account only
a small subset of impurity arrangements. This subset was generated by a
quenching algorithm which imitated vitrification in liquids, by allowing
diffusion of impurity bonds over the lattice. The crossover between the
glassy and high-temperature (melted) state was identified as a supercooled
liquid. To distinguish from customary usage of the term `spin glass', and to
emphasize our glass is kinetic, not thermodynamic, we propose the terms `bond
glass' and `supercooled bond liquid', respectively. A macroscopic criterion
was shown to distinguish the supercooled liquid from the amorphous solid,
connected to the ground-state manifold by slow cooling, which is this model's
analogue of the crystal state of vitreous liquids. The criterion makes it
possible to study field and temperature hysteretic effects on the formation
of the supercooled bond liquid.

\section{Acknowledgements}

Conversations with K.~Uzelac and S.~Bari\v si\'c, and one with M.~M\'ezard,
are gratefully acknowledged. This work was supported by the Croatian
Government under Project $119\,204$.

\end{document}